\documentstyle[aps,prl,graphicx,floats,12pt]{revtex} 
\newif\ifoldrevtex \oldrevtextrue
\newif\iftwocolumns \twocolumnsfalse

\ifoldrevtex
 \newcommand{\affiliation}{\address}
 \newcommand{\email}[1]{\relax}
\fi

\renewcommand{\apj}[3]{Astrophys.\ J.\ {\bf #1}, #3 (#2)}
\renewcommand{\prl}[3]{Phys.\ Rev.\ Lett. {\bf #1}, #3 (#2)}
\renewcommand{\prd}[3]{Phys.\ Rev.\ {\bf D#1}, #3 (#2)}

\newcommand{\rhom}{\rho_M}
\newcommand{\rhomo}{\rho_{M0}}
\newcommand{\rhocard}{\rho_{\rm Card}}
\newcommand{\rhoint}{\rho_{\rm internal}}
\newcommand{\nix}{{\phantom{M}}}
\newcommand{\card}{{\rm Card}}

\begin{document}

\ifoldrevtex\iftwocolumns\twocolumn[\hsize\textwidth\columnwidth\hsize\csname
@twocolumnfalse\endcsname\fi\fi


\title{Fluid Interpretation of Cardassian Expansion}

\author{Paolo Gondolo} 
\email{pxg26@po.cwru.edu} 
\affiliation{Case Western Reserve University, Department of Physics, 10900
  Euclid Ave, Cleveland, OH 44106-7079}

\author{Katherine Freese}
\email{ktfreese@umich.edu}
\affiliation{Michigan Center for Theoretical Physics, University of Michigan,
  Ann Arbor, MI 48109, USA}

\ifoldrevtex\maketitle\fi

\begin{abstract}
  
  A fluid interpretation of Cardassian expansion is developed.  
Here, the Friedmann
equation takes the form $H^2 = g(\rho_M)$ where $\rho_M$ contains only
matter and radiation (no vacuum).  The function $g(\rhom)$ returns to
the usual $8\pi\rhom/(3 m_{pl}^2)$ during the early history of the
universe, but takes a different form that drives an accelerated
expansion after a redshift $z \sim 1$.  One possible interpretation of
this function (and of the right hand side of Einstein's equations) is
that it describes a fluid with total energy density $\rho_{tot} = {3
m_{pl}^2 \over 8 \pi} g(\rhom) = \rhom + \rho_K$ containing not only
matter density (mass times number density) but also interaction terms
$\rho_K$. These interaction terms give rise to an effective negative
pressure which drives cosmological acceleration.  These interactions
may be due to interacting dark matter, e.g. with a fifth force between
particles $F \sim r^{\alpha -1}$. Such interactions may be
intrinsically four dimensional or may result from higher dimensional
physics.  A fully relativistic fluid model is developed here, with
conservation of energy, momentum, and particle number.  A modified
Poisson's equation is derived.  A study of fluctuations in the early
universe is presented, although a fully relativistic treatment of the
perturbations including gauge choice is as yet incomplete.

\end{abstract}

\pacs{}

\ifoldrevtex\iftwocolumns\vskip2.0pc]\fi\else\maketitle\fi


\section{Introduction}

Recent observations of Type IA Supernovae \cite{SN1,SN2}, as well as
concordance with other observations, including the microwave background
\cite{boom} and galaxy power spectra \cite{2df}, indicate that the universe is
flat and accelerating.  Many authors have explored possible explanations for
the acceleration: a cosmological constant, time-dependent vacuum energy such
as quintessence \cite{fafm,peebrat,frieman,stein,caldwell,huey},
and gravitational leakage into extra dimensions \cite{ddg}.

Recently, Freese and Lewis \cite{freeselewis} (Paper I) proposed an explanation
for the acceleration which involves only matter and radiation, invoking no
vacuum energy or cosmological constant whatsoever.  In their model, called
Cardassian, the universe has a flat geometry as required by measurements of the
cosmic background radiation \cite{boom} and yet consists only of matter
and radiation.  The Friedmann equation is modified from its usual form, $H^2 =
{8 \pi \over 3 m_{pl}^2} \rho_M$, to
\begin{equation}
\label{eq:newfunc}
H^2 = g(\rho_M) ,
\end{equation}
where $H = \dot a / a$ is the Hubble constant (as a function of time),
$a$ is the scale factor of the universe, and the energy density
$\rhom$ contains only ordinary matter and radiation.  The function
$g(\rhom)$ reduces to ${8\pi \over 3 m_{pl}^2}\rhom$ in the early
universe, so that Eq.~(\ref{eq:newfunc}) reduces to the ordinary
Friedmann equation during early epochs including primordial
nucleosynthesis.  Only at redshifts $z<{\cal O}(1)$ does the function
$g(\rhom)$ differ from the ordinary Friedmann Robertson Walker (FRW)
case; during these late epochs, $g(\rhom)$ gives rise to accelerated
expansion.  
In Paper I,
the specific form of $g(\rho)$ that was considered was Power Law 
Cardassian,
\begin{equation}
\label{eq:new}
H^2 = {8\pi\over 3 m_{pl}^2} \rhom + B \rhom^n ,
\end{equation}
with
\begin{equation}
n< 2/3.
\end{equation}
The second term only becomes important once
$z<{\cal O}(1)$, at which point it dominates the equation and
causes the universe to accelerate.  Other
possible functions $g(\rho)$ \cite{freese} are discussed further below.

There remains the question of the fundamental origin of these
modifications to the Friedmann equation.  There is no unique
four-dimensional or even higher-dimensional theory that gives
Cardassian evolution.  We consider two different motivations for these
modifications: \hfill\break 
1) These functions may arise from
fundamental theories of gravity in higher dimensions, as was discussed
in \cite{freeselewis}.  Chung and Freese \cite{chung} showed that,
generically, the Friedmann equations are modified as a consequence of
embedding our universe as a three-dimensional surface (3-brane) in
higher dimensions. \hfill\break 
2) Alternatively these functions may arise in a purely
four-dimensional theory in which the modified right hand side of the
Friedmann equation is due to an extra contribution to the total energy
density.  The right hand side is treated as a single fluid, with an
extra contribution to the energy-momentum tensor in (ordinary four
dimensional) Einstein's equations. \hfill\break
The two motivations may or may not be linked, in that the fluid
interpretation may be intrinsically four-dimensional, or it may be an
effective description of higher dimensional physics.

In this paper, we restrict our discussion to four dimensions, and
treat the right hand side of Einstein's equations as a single fluid.
We consider models with an extra energy density associated with matter
that contributes in such a way as to drive acceleration.  This extra
energy density may be intrinsically four dimensional or may serve as
an effective description of higher dimensional physics.  We take the
total energy density of the matter
\begin{equation}
\rho_{tot} = {3 m_{pl}^2 \over 8\pi} g(\rho) = \rho_M + \rho_K
\end{equation}
(plus possible internal thermal energy which is unimportant on
cosmological scales) 
to contain not only the ordinary mass
density $\rho_M$ (mass times number density) but also an additional
contribution $\rho_K$. For example, in Eq.~(\ref{eq:new}), 
\begin{equation}
\rho_K = \frac{3
m_{pl}^2}{8\pi} B \rhom^n.
\end{equation} 

Given this total energy density, we can now compute the
accompanying pressure, and find that the Cardassian contribution
has a {\it negative pressure}, $p_K<0$.  This negative
pressure is responsible for the universe's acceleration.
In fact, one can obtain any negative equation of state
$w_K = p_K/\rho_K <0$, including $w_K<-1$.

The fluid approach has several advantages: (i) it is fully
relativistic, (ii) it allows for the conservation of energy and
momentum as well as of particle number, (iii) it admits a sensible
weak-field limit which leads to a modified Poisson's equation, and
(iv) it permits the study of fluctuations in the early universe, the
study of effects on the cosmic microwave background anisotropies, and
other observables.

The primary purpose of this paper is to examine this fluid approach.
However, we briefly speculate on a possible origin for this extra
term $\rho_K$ in the energy density.  It may arise from (dark)
matter self-interactions that contribute a negative pressure, for
example through a long-range confining force which may be of
gravitational origin or may be a fifth force.  This self-interacting
dark matter is different from any such component considered in the
past, in that is has a negative rather than a positive pressure.  We
speculate on a form of the force between particles that may be
responsible for such an interaction, $F \sim r^{\alpha-1}$, although
this Newtonian form must of course be modified on horizon scales.
This description of a self-interacting dark fluid may be an effective
description of a more fundamental theory.  The fluid approach does
not rely on the validity of such an interpretation of self-interacting
dark matter, e.g., the interactions  may be an effective description
of higher dimensional physics.  

We begin by reviewing the idea of Cardassian expansion in
Sect.~\ref{sec:cardass}.  We present a general fluid formulation in
Sect.~\ref{sec:basics}, and then give specific examples in
Sect.~\ref{sec:examples}. In Sect.~\ref{sec:perturbations} we address the
growth of density perturbations, and in Sect.~\ref{sec:confine} we speculate on
the possible origin of an interaction energy with negative pressure.

\section{Review of Cardassian models}
\label{sec:cardass}

The general form of a Cardassian model was described in
Eq.~(\ref{eq:newfunc}), in which a general function of matter density
replaces the ordinary energy density in the Friedmann equation.  The
simplest version, the Power Law Cardassian model of
Eq.~(\ref{eq:new}), can equivalently be written as
\begin{equation}
  \label{eq:friedcard}
  H^2 = \frac{ 8 \pi G}{3} \, \rho_M \left[ 1 + 
    \left( \frac{\rhom}{\rhocard} \right)^{\!\!n-1\,} \right] .
\end{equation}
For $n<2/3$, the new term is
negligible initially, 
and only comes to dominate at redshift $z \sim 1$
(once $\rhom \sim \rhocard$); once it dominates,
it causes the universe to accelerate. 
We can consider the contribution
of ordinary matter, with
\begin{equation}
\label{eq:matter}
\rho_M = \rho_{M,0} (a/a_0)^{-3}
\end{equation}
to this
new term.  Here, subscript $0$ refers to today.
Once the new term dominates the right hand side of the equation,
we have accelerated expansion.  When the new term is so large
that the ordinary first term can be neglected, the solution
to Eq.~(\ref{eq:new}) is
\begin{equation}
a \propto t^{2 \over 3n}
\end{equation}
so that the expansion is superluminal (accelerated) for $n<2/3$.

The Cardassian model also has the attractive feature that matter alone is
sufficient to provide a flat geometry.  The numerical value of the critical
mass density for which the universe is flat can be modified.  For example, in
Paper I it was shown that in the model of Eq.~(\ref{eq:new}), the value of the
critical mass density can be 0.3 of the usual value.  Hence the matter density
can have exactly this new critical value and satisfy all the observational
constraints such as given by the baryon cluster fraction and the galaxy power
spectrum.

In a `generalized Cardassian model,' other functions $g(\rhom)$ of the matter
(or radiation) density on the right hand side of the Friedmann equation can
also drive an accelerated expansion in the recent past of the universe without
affecting its early history \cite{freese}.  Several of these alternative
functions will be discussed below [see Eqs.~(\ref{eq:polytropic}) and
(\ref{eq:modpoly}) below].

\section{Basic equations}
\label{sec:basics}

\subsection{Perfect Fluid}

We use the ordinary four-dimensional Einstein's equations
\begin{equation}
  \label{eq:einstein}
  G_{\mu\nu} = 8 \pi G T_{\mu\nu} .
\end{equation}
On the right hand side, we take as our ansatz that the energy-momentum
tensor is made only of matter and radiation, and 
has the perfect fluid form,
\begin{equation}
  T^{\mu\nu} = p g^{\mu\nu} + (p+\rho) u^{\mu} u^{\nu} ,
\end{equation}
where $p$ is the pressure, $\rho$ is the {\it total} energy density
of the matter and radiation, and $u^\mu$
is the fluid four-velocity.  
Here the {\it total} energy density for matter includes not only
the mass density $\rhom$ (mass times number density) but also
any interactions or additional terms (corresponding to the new
terms on the right hand side of Eq.~(\ref{eq:newfunc})).
In general, $p$ and $\rho$ are functions of
the mass density $\rhom$ and of thermodynamic variables. 
We assume that at
recent times matter is non-relativistic, 
i.e.\ the typical speeds involved are
much smaller than the speed of light, but we do not assume that $p \ll
\rho$.  

As our ansatz, we take the
total energy
density of the fluid during the matter dominated era to
arise from the sum of three terms: 
\begin{eqnarray}
  \label{eq:split-ord-e}
  \rho & = & \rhom + \rhoint + \rho_{K} , \\
  \label{eq:split-ord-p}
  p & = & p_M + p_{K} , 
\end{eqnarray}
where 
\begin{equation}
  \rhom = m n_M
\end{equation}
is the ordinary matter density of some particle of mass
$m$ and number density $n_M$, $\rhoint$ and $p_M$ are the ordinary
internal energy density and pressure of matter (for example, for an ideal
monoatomic gas at temperature $T_M$, $\rhoint=\frac{3}{2} n_M T_M$ and $p_M=n_M
T_M$), and $\rho_{K}$ and $p_{K}$ are extra (Cardassian) contributions to
energy and pressure.  With regard to the gross properties of the universe (such
as its expansion), we can ignore $\rhoint$ compared to $\rhom$ and $p_M$
compared to $p_K$ for nonrelativistic matter; however, in the context of
galaxies, $\rhoint$ and $p_M$ can be important.

Eq.~(\ref{eq:split-ord-e}) is quite general. Even in those cases where the
total energy density is not a simple linear sum of terms (see Sections IIB and
IIC below), one can always write Eq.~(\ref{eq:split-ord-e}) in this fashion.

In Cardassian models we assume
there is no vacuum contribution to the energy density. We also assume that the
new (Cardassian) contributions $\rho_{K}$ and $p_K$ are only functions of
$\rhom$; i.e., the Cardassian fluid is barotropic.  For example, in Paper I, we
took
\begin{equation}
\rho_K = b \rhom^n,
\end{equation}
where 
\begin{equation}
  b = \frac{ 3 m_{pl}^2 } {8 \pi} B
\end{equation}
with $B$ as in Eq.~(\ref{eq:new}).  

Note that, throughout the rest of this paper, we will focus on the matter
dominated era and hence concentrate on the matter contribution to the fluid
(rather than the radiation).

With the ansatz of Eq.~(\ref{eq:split-ord-e}) and
(\ref{eq:split-ord-p}), Eq.~(1) can be rewritten
\begin{equation}
\label{eq:normal}
H^2 = 8 \pi G \rho/3.
\end{equation}
Hence we recover the ordinary FRW equations, but with
a modified energy density on the right hand side.

One can think of this total energy density as including the effects of an
interaction term: perhaps this term is simply an effective term on large scales
(possibly arising from extra dimensions) describing the expansion of the
universe; or perhaps this term is due to the interaction energy of the dark
matter. A possible origin of an interaction energy with a negative pressure is
described in Sect.~\ref{sec:confine}.
 We iterate again that the fluid approach is only
one of the many ways that Cardassian expansion of Eq.~(1) could result.

\subsection{Conservation laws}

The Bianchi identities guarantee the conservation of energy and momentum,
\begin{equation}
  {T^{\mu\nu}}_{;\nu} = 0 .
\end{equation}
One can follow the evolution of the energy density along each fluid world line
using the general relativistic fluid flow equations. In a comoving frame,
energy-momentum conservation gives the (fully relativistic) energy conservation
and Euler equations \cite{hawking,lyth-stewart}
\begin{eqnarray}
  \dot \rho & = & - \, {u^{\mu}}_{;\mu} \, (\rho + p) ,
  \label{eq:energycons} \\ \dot u_\mu & = & - \frac{ h^\nu_\mu
  p_{;\nu} }{ \rho + p } , \label{eq:Euler}
\end{eqnarray}
where the dot denotes a derivative with respect to comoving time and
the tensor $h_{\mu\nu} = g_{\mu\nu} - u_\mu u_\nu$ projects onto
comoving hypersurfaces.

We impose in addition that mass (or equivalently particle number) is
conserved,
\begin{equation}
( \rhom u^\mu )_{;\mu} = 0 .  
\end{equation}
This will give us the usual dependence of the matter mass density on the scale
factor of the universe.

\subsection{Thermodynamics}

General thermodynamic relations connect $p$ and $\rho$. In particular,
consider
the first law of thermodynamics, $T d(sV) = d(\rho V) + p dV$.
In an adiabatic expansion, $d(sV)=0$.  Together with
mass conservation, $d(\rhom V) = 0$, this equation leads to 
\begin{equation}
  \label{eq:p-epsilon-star}
  p = \rhom \left( \frac{ \partial \rho} 
    {\partial \rhom} \right)_{\!\!s\,} -  
  \rho ,
\end{equation}
which allows $p$ to be computed from an expression for $\rho$. It also
allows the speed of sound to be written as
\begin{equation}
  c_s^2 =  \left( \frac{ \partial p} {\partial \rho} \right)_{\!s} =
  \rhom \left( \frac{ \partial^2 \rho} {\partial \rhom^2} \right)_{\!\!s\,} 
  \bigg/  \left( \frac{ \partial \rho} {\partial \rhom} \right)_{\!\!s} .
\end{equation}
We can define $w=p/\rho$, which in general will not be constant.

Using the ansatz given in Eqs.~(\ref{eq:split-ord-e}) 
and (\ref{eq:split-ord-p})
in Eq.~(\ref{eq:p-epsilon-star}), we can now relate the Cardassian
contributions to energy and pressure via
\begin{equation}
  \label{eq:p-epsilon}
  p_{K} = \rhom \left( \frac{ \partial \rho_{K}} 
    {\partial \rhom} \right)_{\!\!s\,} -  
  \rho_{K} .
\end{equation}
Again we can define $w_K = p_K / \rho_K$, which in general will differ from
$w=p/\rho$ and will not be a constant.

\subsection{Newtonian limit}

Now we obtain the basic equations in the Newtonian limit.
In Minkowski space, we write $u^\alpha = \gamma(1,\vec{v})$
and the metric $\eta^{\alpha\beta} = {\rm diag}(-1,1,1,1)$, where
$\gamma = 1/\sqrt{1-v^2}$ and $\vec{v}$ is the 3-dimensional fluid velocity. 
Then, $T_{\mu\nu} =
{\rm diag}(-\rho,p,p,p)$.

We can obtain the gravitational field equations.
In a weak static field produced by a nonrelativistic mass density,
the time-time component of the metric tensor is approximately given by
$g_{00} \simeq -(1 + 2\phi)$.  Here $\phi$ is the Newtonian potential
for the gravitational field ${\vec{g}} = - \vec{\nabla} \phi$.
The Ricci scalar $ R \simeq G_{00} \simeq 2 \nabla^2
\phi$. Poisson's equation follows from $R=8\pi G {T^\mu}_\mu$ as
\begin{equation}
  \label{eq:poisson}
  \nabla^2 \phi = 4 \pi G (\rho + 3 p ) . 
\end{equation}
We also have
\begin{equation}
\label{eq:curl}
\vec{\nabla} \times \vec{g} =0.
\end{equation}

In the same weak field limit, and for a fluid moving non-relativistically,
the energy conservation and Euler's equations can be found following
Ref.~\cite{weinberg}, but without assuming $p \ll \rho$.
From ${T^{0\beta}}_{;\beta} =0$
we find
\begin{equation}
\label{eq:cont}
{\partial{\rho} \over \partial{t}} + \vec{\nabla} \cdot [(\rho 
+ p ) \vec{v}] = 0.
\end{equation}
From ${T^{i\beta}}_{;\beta} = 0$, we find Euler's equation,
\begin{equation}
\label{eq:nonrel-euler}
{\partial{\vec{v}} \over \partial{t}} + (\vec{v} \cdot \vec{\nabla})
\vec{v} = -  {\vec{\nabla}p \over \rho+ p} - \vec{\nabla}\phi.
\end{equation}
Notice that we do not assume $p \ll \rho$ in the right hand side of
Eqs.~(\ref{eq:poisson}-\ref{eq:nonrel-euler}).

With the additional constraint of particle number conservation,
we have the ordinary continuity equation for matter,
\begin{equation}
\label{eq:numbercons}
{\partial \rho_M \over \partial t} + \vec{\nabla} \cdot (\rho_M
\vec{v}) = 0 .
\end{equation}
Using this equation, 
and in those situations where the internal energy and pressure can
be neglected (e.g.\ when considering the overall expansion of the universe),
we can rewrite Eq.~(\ref{eq:cont}) in the
following way:
\begin{equation}
\label{eq:energy}
\vec{v} \cdot \vec{\nabla}p_K + 
\left( \rho_K + p_K - \rhom {\partial \rho_K \over
\partial \rhom} \right) \vec{\nabla} \cdot \vec{v} = 0 .
\end{equation}
Note that Eq.~(\ref{eq:energy}) in the homogeneous background of our
universe reproduces Eq.~(\ref{eq:p-epsilon}), here in the Newtonian
limit.

Hence our basic nonrelativistic equations are
Eqs.~(\ref{eq:poisson}--\ref{eq:numbercons}).

\subsection{Friedmann-Robertson-Walker models}

As mentioned above,
Friedmann-Robertson-Walker cosmological models take the usual form
\cite{weinberg}. The scale factor of the universe $a$ obeys the equation
\begin{equation}
\label{eq:accel}
  {\ddot a \over a} = - \frac{4 \pi G}{3} (\rho + 3 p)
\end{equation}
and the Friedmann equation
\begin{equation}
  H^2 = \frac{8 \pi G}{3} \, \rho  - \frac{k}{a^2} ,
\end{equation}
where $k=0,\pm1$ fixes the curvature of the spatial sections and $H=\dot a/a$
is the Hubble parameter. 
As mentioned previously, we consider only a flat universe with $k=0$,
as motivated by microwave background data \cite{boom}.
Of course, the total energy density here contains new terms
and is given by Eq.~(\ref{eq:split-ord-e}).

Cardassian expansion requires that the universe be accelerating today.
From Eq.~(\ref{eq:accel}), at matter densities of the order of the
matter density in the universe today, $\rhom \approx \rhomo$, we
require $\rho_0 = \rho(\rhomo)$ and $p_0 = p(\rhomo)$ to satisfy
\begin{equation}
  \label{eq:cond-acc}
  \rho_0 + 3 p_0 < 0 
\end{equation}
so as to have an accelerating universe.  With $p_K$ given in
Eq.~(\ref{eq:p-epsilon}), one can see that acceleration results if $\rho_M -
2 \rho_K + 3 \rho_M {\partial \rho_K \over \partial \rho_M} < 0$.  In
the limit where $\rho_K \gg \rho_M$, one can see that acceleration
results as long as $\rho_K$ goes to zero faster than $\rho_M^{2/3}$
(for $\rho_K =b\rho_M^n$, this requirement becomes $n<2/3$ as stated
previously).

The energy conservation equation in an FRW model is
\begin{equation}
  \dot \rho = - 3 H (\rho + p ) .
\end{equation}
Particle number conservation gives
\begin{equation}
  \dot \rhom = - 3 H \rhom ,
\end{equation}
from which we obtain the usual scaling $ \rhom \sim a^{-3}$
that was used in Eq.~(\ref{eq:matter}).


\section{Examples}
\label{sec:examples}

In this section we discuss three different examples of barotropic
Cardassian models, where the new contribution to the energy density is
a function only of the mass density (mass times number density), $\rho_K
\equiv \rho_K(\rhom)$.  In another paper we examine the consequences for
supernova data of these three models \cite{wang}.

\subsection{Original Power Law Cardassian Model}

In the original Cardassian model of Ref.~\cite{freeselewis},
\begin{equation}
  \rho_{K} = b \rhom^n =
\rhom \left( \frac{\rhocard}{\rhom} \right)^{\!\!1-n\,},
\end{equation}
with $n<2/3$ as in Eqs.~(\ref{eq:new}) and (\ref{eq:friedcard}). 
The pressure associated with this model in the fluid approach follows
from Eq.~(\ref{eq:p-epsilon}) as
\begin{equation}
  \label{eq:pprime2}
  p_{K} = - (1-n) \, \rhom
  \left( \frac{\rhocard}{\rhom} \right)^{\!\!1-n\,} .
\end{equation}
Notice that 
\begin{equation}
  \label{eq:Keos}
  p_{K} = -(1-n) \rho_{K}.
\end{equation}
This model therefore has a constant negative $ w_{K} = p_{K}/\rho_{K} = -(1-n) $.
Then
\begin{equation}
  \label{eq:sum}
  p_K + \rho_K = n \left( \frac{\rhocard}{\rhom}
  \right)^{\!\!1-n\,} \rhom .
\end{equation}
The speed of sound
in this model
\begin{equation}
  c_s^2 = - \frac{ n(1-n) } { n + \left( \frac{\rhom}{\rhocard}
  \right)^{\!\!1-n\,} }
\end{equation}
is not guaranteed to be positive. So this model should be considered as an
effective description at scales where $c_s^2>0$.

\subsubsection{Basic Equations in the Newtonian Limit
for Power Law Cardassian Model}

For the $\rho^n$ Cardassian model, the basic Newtonian equations become:
\begin{equation}
\label{eq:euler2}
{\partial{\vec{v}} \over \partial{t}} + (\vec{v} \cdot \vec{\nabla})
\vec{v} = - {\vec{\nabla}p \over \rho +  p} + \vec{g} ,
\end{equation}
\begin{equation}
\label{eq:energy2}
\vec{v} \cdot \vec{\nabla} p_K+ \left[p_K+ (1-n) \left({\rhocard \over
\rhom}\right)^{1-n} \rho_M\right] \vec{\nabla} \cdot \vec{v} = 0 ,
\end{equation}
\begin{equation}
\vec{\nabla} \times \vec{g} =0,
\end{equation}
\begin{equation}
\label{eq:poisson2}
\vec{\nabla} \cdot \vec{g} = - 4 \pi G \left[\rho_M - (2-3n) \left({\rhocard
\over \rhom}\right)^{1-n} 
\rho_M\right] ,
\end{equation}
\begin{equation}
\label{eq:cont2}
{\partial \rho_M \over \partial t} + \vec{\nabla} \cdot (\rho_M \vec{v}) = 0 .
\end{equation}

\subsubsection{Problem on Galactic Scales}

The fluid approach to the $\rho^n$ model cannot be used on galactic
scales.  From Eq.~(\ref{eq:euler2}) and (\ref{eq:sum}), we see that 
\begin{equation}
\label{eq:gal}
{d \vec{v} \over dt} = - {\vec{\nabla} p_M + \vec{\nabla} p_K
\over \rho_M \left[1+ n (\rhocard/\rhom)^{1-n} \right] } +\vec{g}
\end{equation}
where we have dropped $p_M \ll \rhom$ and $\rhoint \ll \rhom$ in the
denominator.  In standard cosmology, one would have 
\begin{equation}
  {d \vec{v} \over dt} = -{
  \vec{\nabla} p_M \over \rho_M } +\vec{g} ,
\end{equation}
where $p_M$ is quite small and gravity dominates.  Here, the fluid version of
the $\rho^n$ Cardassian cosmology has an additional term in both numerator and
denominator.  Since $\rhocard \ll \rhom$ throughout most of the interiors of
galaxies, we can ignore the second term in the denominator.  However, the
second term in the numerator can have drastic effects.  Let us consider, e.g.,
the position of the Sun in the Milky Way.  The local energy density of matter
is roughly $\rhom \approx 10^4 \, \rhocard$. Hence we have $\rhom \gg \rho_K$,
so that one might expect to be able to ignore any effects of Cardassian terms.
However, these expectations are not true, because of the Cardassian pressure,
which in this case dominates: $|p_K| \gg p_M$.  Hence there is a new force that
acts due to the second term in the numerator of Eq.~(\ref{eq:gal}),
\begin{equation}
\left. \frac{d\vec{v}}{dt} \right|_{\rm new}
= -{\vec{\nabla}p_K \over \rho_M} = + n(1-n) \left({\rhocard
\over \rhom} \right)^{1-n}
{\vec{\nabla}\rhom \over \rho_M} .
\end{equation}
This force destroys flat rotation curves (velocities tend to increase
as one goes out to large radii in an unacceptable way).  The argument
in the preceding paragraph has been presented using nonrelativistic
equations, and as such is incomplete.  However, we have found that the result
that a problematic new force arises on galactic scales remains true in
a fully relativistic treatment based on the Oppenheimer-Volkov equation.

The fluid $\rho^n$ Cardassian case must therefore be thought
of as an effective model, which applies only on cosmological scales.
The examples in the following two subsections, on the contrary, can be
treated as fluid models on all scales, including cosmological scales
as well as galactic scales.

\subsection{Polytropic Cardassian}

Another class of 
models has
\begin{equation}
\label{eq:polytropic}
  \rho = \rhoint + 
  \rhocard \left[ 1 + \left( \frac{\rhom}{\rhocard} 
    \right)^{\!\!q\,}\right]^{ \frac{1}{q} }
\end{equation}
with $q \ne 0$.  This model can be used on all scales (see below), but it
does not quite fit the criteria of Cardassian as defined in
Eq.~(\ref{eq:newfunc}); at late times in the future of the universe, when $\rhom
\ll \rhocard$, this model becomes cosmological constant dominated with $\Lambda
= \rhocard$.  Phenomenologically, this energy density is very similar to a
model that was derived earlier \cite{ddg} motivated by gravitational leakage
into extra dimensions.

In this model, the pressure is
\begin{equation}
  p = p_M - \rhocard \left[ 1 + \left( \frac{\rhom}{\rhocard} 
    \right)^{\!\!q\,}\right]^{ \frac{1}{q} - 1} .
\end{equation}
When the ordinary internal energy density $\rhoint$ and the ordinary pressure
$p_M$ can be neglected, this model obeys a polytropic equation of state
\begin{equation}
  p = - \rhocard \left( \frac{\rho}{\rhocard} \right) ^ {1-q} ,
\end{equation}
with negative pressure and negative polytropic index $N=-1/q$. For $q
> 1$, the speed of sound in this model is positive,
\begin{equation}
  c_s^2 = \frac{ q-1 } { 1 + \left( \frac{\rhom}{\rhocard} 
    \right)^{\!\!q\,} } .
\end{equation}
Thus this fluid model can be used on all scales.

We must make sure that at the scales of galaxies and galaxy clusters the
Cardassian pressure can be neglected compared to the ordinary pressure.  At
large matter densities, the Cardassian pressure is $ |p_{K}| \simeq
\rhom^{1-q} \rhocard^{q} $. In a galaxy or cluster with velocity
dispersion $\sigma$, the ordinary pressure is $p_M \simeq \rhom \sigma^2$. We
want $ | p_K | / p_M \simeq (\rhocard/\rhom) ^ {q}/\sigma^2 \ll 1$.
Taking $\sigma \simeq 300 $ km/s and assuming $p_K$ is unimportant out to
$\approx$ 100 kpc where $ \rhom \approx 10^2 \, \rhocard$, this condition
amounts to $ q \gtrsim 3$. It is remarkable that this value of $q$ is
compatible with supernova data \cite{wang}.

\subsection{Modified polytropic Cardassian}

A Cardassian model that can be used on all scales is
\begin{equation}
\label{eq:modpoly}
  \rho = \rhoint + \rhom \left[ 1 + \left(
  \frac{\rhocard}{\rhom} \right)^{\!\!q\nu\,}\right]^{
  \frac{1}{q} }.
\end{equation}
For $q=1$ this reduces to the original $\rho^n$ Cardassian model with
$n=1-\nu$. 
The pressure follows as
\begin{equation}
  p = p_M - \nu \rhom \left[ 1 + \left(
    \frac{\rhocard}{\rhom} \right)^{\!\!q\nu\,}\right]^{
    \frac{1}{q} - 1} \left( \frac{\rhocard}{\rhom}
    \right)^{\!\!q\nu\,} .
\end{equation}
This model is interesting because the two parameters $\nu$ and $q$ are
important on different scales.  The parameter $\nu$ sets the current value of
$w \simeq - \nu$, and so can be chosen to fit the supernova data, while the
parameter $q$ governs the suppression of the Cardassian pressure at high
densities, and can therefore be chosen not to interfere with galactic rotation
curves and cluster dynamics. Concrete comparisons with data will be presented
in \cite{wang}.

\section{Perturbations}
\label{sec:perturbations}

\subsection{Newtonian theory}

Here we calculate the behavior of small fluctuations, using the
nonrelativistic equations derived above in Eqs.~(\ref{eq:poisson} -
\ref{eq:numbercons}).  As discussed below, there are gauge choices
that render the Newtonian theory inadequate, yet we can learn from it
nonetheless.  We will write down the general perturbation equations
for any generalized Cardassian model, and then as an example will
solve them in the $\rho^n$ form of the Cardassian equations (of Paper
I) to illustrate the type of results that occur.


For the zero-order solution (superscript ${}^{(0)}$) we take the simple
spatially uniform solution with
\begin{eqnarray}
\rho_M^{(0)} & = & \rho_{M,0} \left[ {a_0 \over a(t)} \right]^3
\\
\vec{v}^{(0)} & = & {\dot a(t) \over a(t)}  \,\, \vec{r} 
\\
\vec{g}^{(0)} & = & - {4\pi G (\rho^{(0)} + 3 p^{(0)}) \over 3} \,\, \vec{r}.
\end{eqnarray}

We now seek a perturbed solution (superscript ${}^{(1)}$) by adding to the
zero-order solution the small perturbations $\rho_M^{(1)}$,
$\vec{v}_\nix^{(1)}$, $p_M^{(1)}$ and $\vec{g}_\nix^{(1)}$.  Following
\cite{weinberg}, we find that the hydrodynamic equations
Eq.~(\ref{eq:euler2}-\ref{eq:pprime2}) then give, to first order in these
perturbations,
\begin{eqnarray}
\label{eq:pertv}
&& \dot{\vec{v}}{}_\nix^{(1)} + {\dot{a} \over a} \, \vec{v}{}^{(1)} = 
- {1 \over \rho_M + \rho_K} \left(\vec{\nabla} p_{M}^{(1)} + \vec{\nabla} p_{K}^{(1)}\right) +
\vec{g}_\nix^{(1)} ,
\\
&& \dot \rho_{M}{}^{(1)} + 3 \, {\dot a \over a} \, \rho_{M}^{(1)} +
\rho_M \vec{\nabla} \cdot \vec{v}_\nix^{(1)} = 0,
\\
&& \strut \vec{\nabla} \times \vec{g}_\nix^{(1)} = 0,
\\
\label{eq:pertg}
&& \vec{\nabla} \cdot \vec{g}_\nix^{(1)} = - 4 \pi G \left[ 1+ {\partial \over
\partial\rhom}(\rho_K + 3 p_K) \right] \rho_{M}^{(1)} .
\end{eqnarray}
We have dropped the superscript (0) from zero-order quantities. In deriving
these equations we have neglected $\rhoint$ and $p_M$ with respect to $\rhom$.

We take the perturbations to be adiabatic, so that the ordinary pressure
perturbation is given by
\begin{equation}
p_{M}^{(1)} = v_s^2 \, \rho_{M} ^{(1)}
\end{equation}
where $v_s$ is the ordinary speed of sound.  In the generalized Cardassian
model with pressure and energy density related by Eq.~(\ref{eq:p-epsilon}), the
perturbation equations require
\begin{equation}
\label{eq:pertp}
p_{K}^{(1)} = \rhom \frac{\partial ^2 \!\rho_K}{\partial
    \rhom^2}  \,\,
\rho_{M}^{(1)} .
\end{equation}

\subsubsection{Newtonian perturbation equations for $\rho^n$ Cardassian
case}

For the case of Paper I in Eq.~(\ref{eq:new}),
Eqs.~(\ref{eq:pertp}) and (\ref{eq:pertg}) become
\begin{equation}
\vec{\nabla} \cdot \vec{g}_\nix^{(1)} = - 4 \pi G \left[
 1 + n (3n-2) b \rho_M^{n-1} \right]
\rho_{M}^{(1)}
\end{equation}
and
\begin{equation}
p_{M}^{(1)} = b n (n-1) \rho_M^{n-1} \rho_{M}^{(1)} .
\end{equation}

\subsection{Newtonian theory: non-expanding case}

First, let us consider the nonexpanding case with unperturbed velocity $\vec{v}
= ({\dot a/a}) \vec{r} = 0$.  A combination of the perturbation equations
(\ref{eq:pertv}--\ref{eq:pertg}) yields 
\begin{equation}
{\partial^2 \delta_M \over \partial t^2} = v_{s,\rm new}^2 
\, \nabla^2 \delta_M 
+ 4 \pi G \rho_{M,\rm new} \, \delta_M .
\end{equation}
This is the usual perturbation equation for 
\begin{equation}
\delta_M = \frac{ \rho_M^{(1)} } {\rho_M} ,
\end{equation}
with $v_s^2$ replaced by 
\begin{equation}
\label{eq:vsnewgen}
v_{s,\rm new}^2 = \frac{\rho_M}{\rho_M+\rho_K} \left( 
v_s^2 + \rhom {\partial^2\!\rho_K \over \partial \rhom^2} \right)
\end{equation}
and $\rho_M$ replaced by
\begin{equation}
  \label{eq:rhomnewgen}
  \rho_{M,\rm new} = \rhom {\partial \over \partial\rhom} (\rho + 3 p) 
  = \rho_M + \rho_M {\partial \over
\partial\rhom}(\rho_K + 3 p_K) .
\end{equation}

The equations are spatially homogeneous, so we expect to
find plane-wave solutions.  We take
$\delta_M \propto {\rm exp}[i(\vec{k} \cdot \vec{x} - \omega t)]$
to find
\begin{equation}
\omega^2 = k^2 v_{s,\rm new}^2 - 4 \pi G \rho_{M,\rm new}  .
\end{equation}
If $\omega^2 < 0$, then the perturbation is unstable to collapse.
Hence we find the Jeans length
\begin{equation}
\lambda_J = \frac{2\pi}{k_J} = \sqrt{\pi} {v_{s,\rm new} \over \sqrt{G
    \rho_{M,\rm new} } } .
\end{equation}
Perturbations on scales larger than the Jeans length can
collapse.

In particular for the $\rho^n$ Cardassian case of Paper I
in Eq.~(\ref{eq:new}), we have
\begin{equation}
  \label{eq:vsnew}
  v_{s,\rm new}^2 = {v_s^2 - b n(1-n) \rho_M^{n-1}
    \over 1 + n b \rho_M ^{n-1}} .
\end{equation}
and
\begin{equation}
  \label{eq:rhomnew}
  \rho_{M,\rm new} = \rho_M \left[ 1 - b n (2-3n) \rho_M^{n-1} \right].
\end{equation}
Then the Jeans length becomes
\begin{equation}
  \lambda_J = \sqrt{ \frac{ \pi} { G \rho_M } \, \, 
    \frac{ v_s^2 - b n(1-n) \rho_M^{n-1}} { \left[ 1 + n b \rho_M ^{n-1}
      \right] \left[ 1 - b n (2-3n) \rho_M^{n-1} \right] } } .
\end{equation}
The Jeans length for ordinary cold dark matter was tiny,
and the modifications here do not change it substantially
enough to make it interesting.

\subsection{Newtonian theory: expanding case}

Here we consider the expanding case with $\vec{v} = 
({\dot{a} / a }) \vec{r} \neq 0$.
We take the plane wave form for the solutions,
\begin{eqnarray}
&& \rho_{M}^{(1)}(\vec{r},t) = \rho_M(t) \, \delta(\vec{r},t) ,
\\
&& \delta(\vec{r},t) = \delta(t) \,
\exp\!\left[{i\vec{r} \cdot \vec{q} \over a(t)}\right].
\end{eqnarray}

After some algebra, we find
\begin{equation}
\label{eq:ddot}
\ddot{\delta} + {2 \dot{a} \over a} \dot{\delta}
+ \left( v_{s,\rm new}^2 {q^2 \over a^2} - 4 \pi G \rho_{M,\rm new} \right) 
\delta = 0 ,
\end{equation}
where $v_{s,\rm new}^2$ and $\rho_{M,\rm new}$ are given in
Eqs.~(\ref{eq:vsnewgen}) and~(\ref{eq:rhomnewgen}).

\begin{figure}
  \begin{center}
    \includegraphics[width=0.8\textwidth]{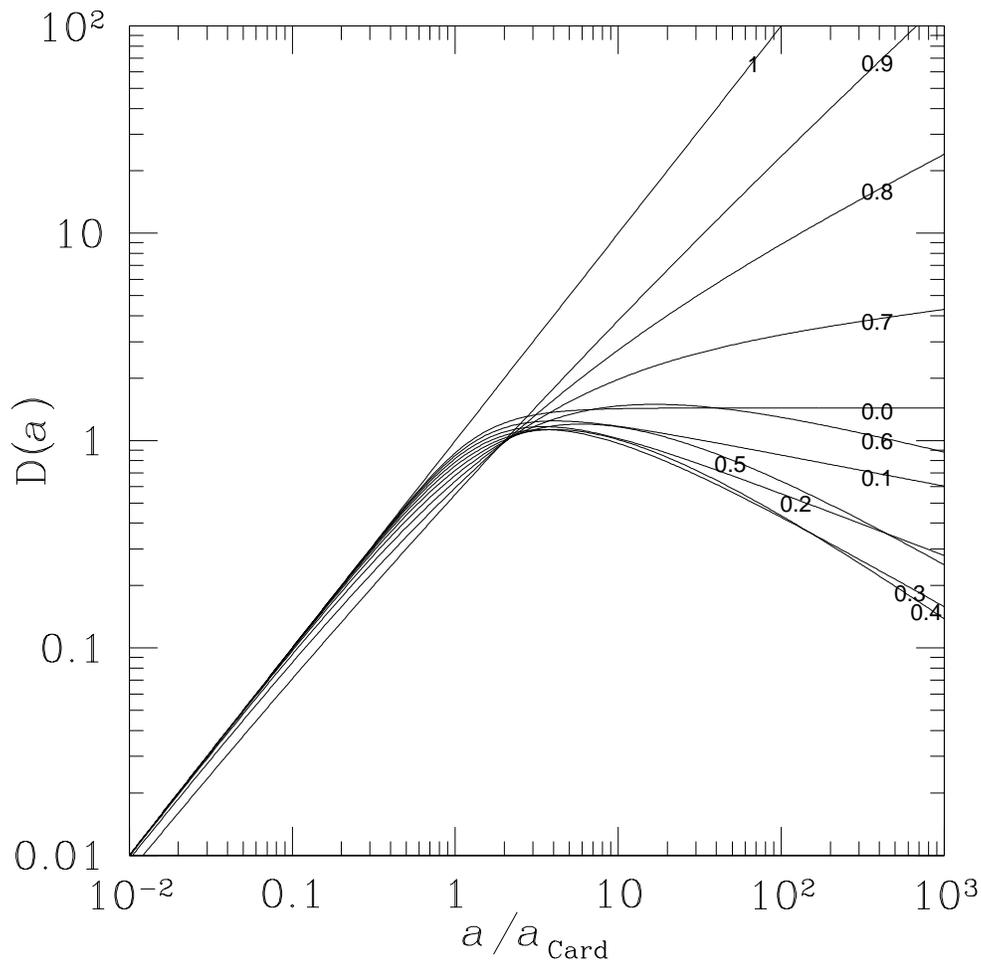}
    \vspace{\baselineskip}
    \caption{
      Growth factor $D(a)$ for matter density perturbations in the original
      $\rho_K = b \rho_M^n$ Cardassian model, as a function of the scale factor
      $a$.  The normalization scale factor $a_\card$ is defined by 
      $\rho_M(a_\card) = \rho_K(a_\card)$. Curves are labeled by the Cardassian
      index $n$. The curve with $n=1$ corresponds to the usual growth in a
      matter dominated universe, while $n=0$ corresponds to a cosmological
      constant.}
  \end{center}
\end{figure}

For the Power Law 
$\rho^n$ Cardassian model of Eq.~(\ref{eq:new}), which is an effective
model on large scales, we can drop the $q^2$ term. We have solved the resulting
equation numerically after changing independent variable from $t$ to $a(t)$.
The solution for $\delta(a)$ (for modes with $\lambda \gg \lambda_J$) is given
in Figure 1 for varying values of the Cardassian index $n$.  In the figure,
$a_\card$ is defined to be the scale factor at which the Cardassian term starts
to dominate in the Friedmann equation, i.e. when $\rho_M(a_\card) =
\rho_K(a_\card)$.  Today, $a_0/a_\card$ is a factor of a few, but the figure
extends to much higher values of $a$ to show the behavior of the solutions. The
vertical axis is normalized so that the solution becomes $a/a_\card$ at early
times; namely, using the solution in Ref.~\cite{groth-peebles} for the growth
of matter fluctuations in the radiation and matter dominated eras of a
cosmologically flat model,
\begin{equation}
  D(a) = \frac{\delta(a)}{\delta_{\rm init}} \frac{ 2 a_{\rm eq} }{ 3 a_\card
  },
\end{equation}
where $\delta_{\rm init}$ is the initial value of the pertubation and $a_{\rm
  eq}$ is the scale factor at matter-radiation equality.  One can see in Figure
1 that perturbation growth is suppressed once $a>a_\card$ for $0 \leq n < 1$.
As a reminder, $n = 1$ corresponds to ordinary matter whereas $n=0$ corresponds
to a cosmological constant.  To remind the reader, this figure corresponds to
perturbation growth in an expanding Newtonian fluid in a box with the sides
being pulled out.

\subsection{General-relativistic theory}

Since we do not neglect the pressure $p$ with respect to the energy
density $\rho$, the perturbations should in fact be studied in a
general-relativistic theory. The problem arises of the choice of
gauge. In order to define the energy density perturbation
throughout spacetime, one needs to choose a set of hypersurfaces;
i.e., one needs to choose a gauge.  Two choices are common:\hfil\break
(1) The comoving gauge corresponds
to the set of comoving hypersurfaces, defined as those which
are orthogonal to the comoving world lines, i.e., to the 
world lines which follow the flow of energy.\hfil\break
(2) The synchronous gauge corresponds to the hypersurfaces which
are orthogonal to geodesics\footnote{There are in fact
an infinity of synchronous gauges, but authors typically drop
the so-called ``gauge mode'' solution so that the synchronous
gauge becomes unique.}.

The value of density
perturbations depends on the choice of gauge.  However,
typically, if one looks inside the horizon and well into 
the matter dominated era, the value of the density perturbation
becomes the same in all gauges.

In the case of Cardassian cosmology, we 
have the unusual circumstance that, even inside
the horizon and well into the matter dominated era, the
value of the density perturbation depends on the choice of gauge.

The evolution of density perturbations in the universe using a fluid flow
approach was discussed in Ref.~\cite{lyth-stewart}.  In the comoving gauge,
with $ w = p/\rho$ and the relation $ \delta p/\delta \rho = c_s^2 $ (true for
barotropic Cardassian models), we find that the fractional perturbation $\delta
= \delta \rho / \rho $ of momentum $k$ obeys the equation~\cite{lyth-stewart}
\begin{equation}
  H^{-2} \ddot \delta + [ 2 - 3(2w-c_s^2)] H^{-1} \dot \delta - 
  \frac{3}{2} ( 1 - 6 c_s^2 + 8 w - 3 w^2) \delta = - \left( \frac{k}{aH}
  \right)^2 c_s^2 \delta .
\end{equation}
The pertubation $\delta_s$ in the synchronous gauge is related to
$\delta$ by \cite{lyth-stewart}
\begin{equation}
  \label{eq:gauges}
  \delta - \delta_s = 3 H (1+w) \int_0^t \frac{ \delta p }{ \rho + p} dt,
\end{equation}
if one drops the ``gauge mode.''  In ordinary cosmologies, where there
is no Cardassian pressure term, the integral on the right hand side
has a fixed value after matter domination, obtained by setting $\delta
p = 0 $ in the matter-dominated era.  A few Hubble times into the
matter-dominated era it becomes negligible compared to the time scales
of interest so that the comoving and synchronous gauges become
identical.  However, the values of the density perturbations in the
two gauges do not become equal in Cardassian cosmology, when the
Cardassian pressure $\delta p_K$ is present. This term contributes to
the integral in Eq.~(\ref{eq:gauges}) all the way to the present time,
and the comoving and synchronous gauges are not identical even
today. This creates a problem of interpretation for fluctuations in
the present universe, which must be addressed in future studies.

\section{Origin of interaction energy with negative pressure}
\label{sec:confine}

Here we speculate on a possible origin for an interaction energy with a
negative pressure.  Dark matter particles may be subject to a new interparticle
force which is long-range and confining,
\begin{equation}
  F(r) \propto r^{\alpha-1}
\end{equation}
with $\alpha>0$. This force may be of gravitational origin or maybe a fifth
force.

To be more quantitative, let us write the new interparticle potential as
\begin{equation}
  U_{ij} = A r_{ij}^\alpha ,
\end{equation}
where $r_{ij}$ is the distance between particles and $A$ is a normalization
constant. The total new interaction energy of a system of $N$ particles
occupying a volume of radius $R$ will be
\begin{equation}
  U_{\rm new} \simeq A N^2 R^\alpha,
\end{equation}
to within a numerical factor of order 1 dependent on the geometry. The
total gravitational potential energy of the same system is, also within a
factor of order unity,
\begin{equation}
  U_{\rm grav} \simeq \frac{ G M^2 }{R} ,
\end{equation}
where $ M$ is the total mass of the system. To play a
cosmological role at the present time, the new energy must be of the same order
of the gravitational energy when $R \simeq R_H$, the current size of the
horizon. Imposing that $U_{\rm new} \simeq U_{\rm grav}$ at $ R \simeq R_H$
gives us the normalization
\begin{equation}
  A = \frac{G m^2}{R_H^{\alpha+1}} ,
\end{equation}
where $m=M/N$ is the mass of a single particle.

We can now find the magnitude of the new force on galactic scales. We have
\begin{equation}
  U_{ij}(r) \simeq \frac{G m^2}{R_H} \left( \frac{r}{R_H} \right)^\alpha .
\end{equation}
Thus the new force per unit mass on a particle of mass $m$ at distance $R_{\rm
  g}$ from a system of $N$ particles is of order
\begin{equation}
  \frac{F_{\rm new}}{m} \simeq \frac{ | \vec{\nabla} U_{\rm new} |}{m} \simeq
  \alpha \,\frac{G M}{R_H^2} \left( \frac{R_{\rm g} }{R_H} \right)^{\alpha-1} .
\end{equation}
Compared with the gravitational force
\begin{equation}
  \frac{F_{\rm grav}}{m} \simeq \frac{G M}{R^2_{\rm g}} ,
\end{equation}
this gives
\begin{equation}
  \frac{ F_{\rm new}}{F_{\rm grav}} \simeq \alpha \left( \frac{ R_{\rm g}}{R_H}
  \right)^{\alpha+1} . 
\end{equation}
For $R_{\rm g}$ of the order of galactic scales, i.e.\ $R_{\rm g} \ll R_H$ the
new force is negligible compared to the gravitational force. It is clear that
for $\alpha>0$ the new force is only important on very large scales.  This
Newtonian formulation must of course be modified at large distances because of
the finite speed of light and issues of causality.

We want to comment on the equation of state of a system subject to long-range
confining forces. If such a system reaches thermal equilibrium (that it does so
in the presence of long-range confining forces is not at all clear), then
simple statistical mechanics considerations based on the scaling of the
partition function lead to the equation of state (see Appendix for details)
\begin{equation}
  \label{eq:pconf}
  p = -\frac{\alpha}{3} \rho .
\end{equation}
This is the equation of state of the force mediators. For example, the Coulomb
force (although not confining) has $\alpha=-1$ and equation of state
$p=\rho/3$, which is that of photons. If $\alpha=1$ (as in QCD) or
$\alpha=2$, the mediators are strings and 2-dimensional objects, respectively,
and their equations of state are $p=-(1/3) \rho$ and $p = -(2/3) \rho$, which
are those of a network of strings and of domain walls, respectively.  Finally,
one obtains the vacuum equation of state $p=-\rho$ for $\alpha=3$. 
Notice that the Cardassian index $n$ in the Power Law
$\rho^n$ model is connected to
the exponent $\alpha$ in the confining force law through $\alpha = 3(1-n)$
(cfr.\ e.g.\ Eqs.~(\ref{eq:Keos}) and (\ref{eq:pconf})). That a
confining force can give rise to an effective negative pressure is well-known
in particle physics, where the MIT bag model is just such an effective
description of quark confinement. The cosmological negative pressure may be an
indication that our observable universe is in a big bag. This suggests that
the dark energy may be the interaction energy associated to a long-range
confining force. 

\section{Conclusions}

An interpretation of Cardassian expansion as an interacting dark matter fluid
with negative pressure is developed. The Cardassian term on the right hand side
of the Friedmann equation (and of Einstein's equations) is interpreted as an
interaction term. So the total energy density contains not only the matter
density (mass times number density) but also interaction terms. These
interaction terms give rise to an effective negative pressure which drives
cosmological acceleration.

These interactions may be due to interacting dark matter, e.g. with a
long-range confining force or a fifth force between particles. Alternatively,
such interactions may be an effective description of higher dimensional
physics.  We have said that matter alone can be responsible for accelerated
behavior.  However, if the Cardassian behavior results from integrating 
out extra dimensions, then one may ask what behavior of the radii of
the extra dimensions is required.  Similarly, if we follow a QCD bag
or other description of self-interacting dark matter, one may wonder
if an equivalent vacuum description can be constructed.  Further work
in search of a fundamental origin of Cardassian expansion must be studied
to answer these questions in detail.

A fully relativistic fluid model of Cardassian expansion has been developed, in
which energy, momentum, and particle number are conserved, the modified
Poisson's equations have been derived, and a preliminary study of density
fluctuations in the early universe has been presented.

One of the goals of this study is to allow predictions of various
observables that will serve as tests of the model.  The Cardassian
model will have unique predictions, particularly due to the modified
Poisson's equations. For example, one can now calculate the effect on
the Integrated Sachs Wolfe effect in the Cosmic Microwave Background
\cite{lef}.  In addition, one can now calculate the effect on cluster
abundances as function of redshift. These predictions can then be
tested against existing and upcoming measurements of these quantities.
Comparison with existing and upcoming supernova data is being studied
in another paper \cite{wang}.  We reiterate that this fluid approach
is only one of the ways that Cardassian expansion may result.

\acknowledgments

We thank Ted Baltz, Richard Easther, Josh Frieman, and Yun Wang for
useful conversations.  K.F.\ acknowledges support from the Department
of Energy via the University of Michigan. This research was supported
in part by the National Science Foundation under Grant No.\
PHY99-07949 at the Kavli Institute for Theoretical Physics, University
of California, Santa Barbara.  K.F. thanks the Aspen Center for
Physics for hospitality during her stay. P.G. thanks the Michigan Center for
Theoretical Physics for hospitality.

\appendix

\section{Equation of state for a system subject to confining forces}

We give here the details of the derivation of the equation of state $ p =
-(\alpha/3) \rho $ for the mediators of a confining interparticle potential $U
= A r^\alpha$. Our derivation assumes that the gas is non-relativistic and in
thermodynamical equilibrium. We stress that it is not at all clear that a
system of particles subject to long-range confining forces may reach thermal
equilibrium. For our application to dark energy, it is not even clear that it
could reach equilibrium on a time scale short compared to the age of the
Universe. Therefore, the content of this Appendix may be of somewhat academic
interest.  Nevertheless, we present it for completeness.

The partition function of a non-relativistc gas of $N$ particles subject to a
confining interparticle potential $U = A r^\alpha$ is
\begin{equation}
  Z(V,T) = \prod_{i=1}^{N} \int_V d^3r_i \int \frac{d^3p_i}{(2\pi)^3} 
\exp\left[{- \sum_j \frac{p_j^2}{2mT} -
  \sum_{j<k} \frac{A r_{jk}^\alpha}{T} } \right] 
\end{equation}
where $T$ is the temperature and $V$ is the volume occupied by the system.

If we rescale $V \to \lambda^3 V$ and $T \to \lambda^\alpha T$, and then change
integration variables $r \to \lambda r'$, $p \to \lambda^{\alpha/2} p'$, we
can prove that the partition function scales as
\begin{equation}
  Z(\lambda^3 V, \lambda^\alpha T) = 
\lambda^{3N+\frac{3}{2}\alpha N} Z(V,T) .
\end{equation}
Now the free (ideal gas) partition function
\begin{equation}
  Z_0(V,T) =     V^N \left( \frac{m T}{2 \pi} \right)^{\frac{3}{2} N} 
\end{equation}
scales in the same way,
\begin{equation}
  Z_0(\lambda^3 V, \lambda^\alpha T) = 
\lambda^{3N+\frac{3}{2}\alpha N} Z_0(V,T) .  
\end{equation}
It follows that 
\begin{equation}
  Z(V,T) = Z_0(V,T) \, Z_1(T^3/V^\alpha),
\end{equation}
where $Z_1(x)$ is a function of the ratio $T^3/V^\alpha$, which is invariant
under the rescaling $V \to \lambda^3 V$, $T \to \lambda^\alpha T$.

Pressure, entropy, and energy density can then be computed from the free energy
\begin{equation}
  F = - T \ln Z(V,T) = - T \ln Z_0(V,T) - T \ln Z_1(T^3/V^\alpha) 
\end{equation}
as
\begin{eqnarray}
  \label{eq:A1}
  P & = &- \left( \frac{ \partial F } { \partial V } \right)_T = 
  \frac{NT}{V} - \frac{ \alpha NT}{V} f(T^3/V^\alpha) , \\
  \label{eq:A2}
  U & = & F - T \left( \frac{ \partial F } { \partial T } \right)_V =
  \frac{3}{2} NT + 3 NT f(T^3/V^\alpha) .
\end{eqnarray}
where
\begin{equation}
f(x) = \frac{1}{N} \frac{x}{Z_1} \frac{dZ_1}{dx} .
\end{equation}
The first terms on the right-hand sides of eqs.~(\ref{eq:A1}) and~(\ref{eq:A2})
correspond to the ideal gas. The second terms are the pressure $p$ and the
energy $\rho V$ due to the confining forces. They obey the relation
\begin{equation}
  \label{eq:A3}
  p = - \frac{\alpha}{3} \rho.
\end{equation}
Although we have not quantized the interaction, we draw on the analogy with
electromagnetism described in the main text and call eq.~(\ref{eq:A3}) the
equation of state for the force mediators.



\begin{thebibliography}{99}
  
\bibitem{SN1} S.~Perlmutter {\it et al.}  [Supernova Cosmology Project
  Collaboration], Astrophys.\ J.\ {\bf 517}, 565 (1999) [astro-ph/9812133].
  
\bibitem{SN2} A.~G.~Riess {\it et al.}  [Supernova Search Team
  Collaboration], Astron.\ J.\ {\bf 116}, 1009 (1998)
  [astro-ph/9805201].
  
\bibitem{boom} C.B.~Netterfield {\it et al}, astro-ph/0104460; R.~Stompor {\it
    et al}, astro-ph/1015062; N.W.~Halverson {\it et al}, astro-ph/0104489; C.
  Pryke {\it et al} \apj{568}{2002}{46}

\bibitem{2df} L.~Verde {\it et al.} [2dF Survey] MNRAS {\bf 335}, 432 (2002)

\bibitem{fafm} K. Freese, F.C. Adams, J.A. Frieman, and E. Mottola,
  Nucl.\ Phys.\ {\bf B287}, 797 (1987).
  
\bibitem{peebrat} P.J.E. Peebles and B. Ratra, Astrophys.\ J.\ Lett.
{\bf 325}, L17 (1988); B. Ratra and P.J.E. Peebles, \prd{37}, 3406 (1988).

\bibitem{frieman} J. Frieman, C. Hill, A. Stebbins, and I. Waga, \prl{75}
  {1995}{2077}.

\bibitem{stein} L. Wang and P. Steinhardt, \apj{508}{1998}{483}.
  
\bibitem{caldwell} R. Caldwell, R. Dave, and P. Steinhardt,
  \prl{80}{1998}{1582}
  
\bibitem{huey} G. Huey, L. Wang, R. Dave, R. Caldwell, and P. Steinhardt,
  \prd{59}{1999}{063005}
  
\bibitem{ddg} C.~Deffayet, Phys.\ Lett.\ B {\bf 502}, 199 (2001);
C.~Deffayet, G.~Dvali, and G.~Gabadadze, Phys.\ Rev.\
  {\bf D65}, 044023 (2002).

\bibitem{freeselewis} K.~Freese and M.~Lewis, Phys.\ Lett.\ {\bf B540}, 1
  (2002) [astro-ph/0201229].

\bibitem{freese} K.~Freese, hep-ph/0208264.

\bibitem{chung} D.J. Chung and K. Freese, \prd{61}{2000}{023511}
  
\bibitem{hawking} S.~Hawking, Ap. \ J. \ {\bf 145}, 544 (1966).

\bibitem{lyth-stewart} D.~Lyth and E.~Stewart, 
Ap. \ J. \ {\bf 361}, 343 (1990).

\bibitem{weinberg} S.~Weinberg, {\it Gravitation and cosmology} (John
  Wiley and Sons, New York, 1972).

\bibitem{wang} Y.~Wang, K.~Freese, P.~Gondolo, and M. Lewis,
astro-ph/0302064.

\bibitem{groth-peebles} E.~J.~Groth and P. J. E. Peebles, Astron.\ Astrophys.\
  {\bf 41}, 143 (1975).

\bibitem{lef} M.~Lewis, R.~Easther, and K.~Freese,
in preparation.

\end{thebibliography}
\end{document}